\documentclass[
 reprint,
superscriptaddress,
 amsmath,amssymb,
 aps,
]{revtex4-2}
\usepackage{xcolor}
\usepackage{graphicx}
\usepackage{dcolumn}
\usepackage{bm}
\usepackage{gensymb}
\usepackage{soul}
\usepackage{lineno}
\usepackage{amsmath}

\begin{document}

\preprint{APS/123-QED}

\title{Electrical manipulation of intervalley trions in twisted MoSe$_2$ homobilayers at room temperature}

\author{Bárbara L. T. Rosa}\email{rosa@physik.tu-berlin.de}
\affiliation{Institute of Solid State Physics, Technische Universität Berlin, 10623 Berlin, Germany}
\affiliation{Institute of Physics “Gleb Wataghin”, State University of Campinas, 13083-859 Campinas, Brazil}

\author{Paulo E. {Faria~Junior}} 
\affiliation{Institute for Theoretical Physics, University of Regensburg, 93040 Regensburg, Germany}

\author{Alisson R. Cadore}
\affiliation{Brazilian Nanotechnology National Laboratory, Brazilian Center for Research in Energy and Materials, Campinas, Sao Paulo 13083-970, Brazil}

\author{Yuhui Yang}
\affiliation{Institute of Solid State Physics, Technische Universität Berlin, 10623 Berlin, Germany}

\author{Aris Koulas-Simos}
\affiliation{Institute of Solid State Physics, Technische Universität Berlin, 10623 Berlin, Germany}

\author{Chirag C. Palekar}
\affiliation{Institute of Solid State Physics, Technische Universität Berlin, 10623 Berlin, Germany}

\author{Sefaattin Tongay}
\affiliation{School for Engineering of Matter, Transport and Energy, Arizona State University, Tempe, USA}

\author{Jaroslav Fabian}
\affiliation{Institute for Theoretical Physics, University of Regensburg, 93040 Regensburg, Germany}


\author{Stephan Reitzenstein}\email{stephan.reitzenstein@physik.tu-berlin.de}
\affiliation{Institute of Solid State Physics, Technische Universität Berlin, 10623 Berlin, Germany}






\begin{abstract}
The impressive physics and applications of intra- and interlayer excitons in a transition metal dichalcogenide twisted-bilayer make these systems compelling platforms for exploring the manipulation of their optoelectronic properties through electrical fields. This work studies the electrical control of excitonic complexes in twisted MoSe$_2$ homobilayer devices at room temperature. Gate-dependent micro-photoluminescence spectroscopy reveals an energy tunability of several meVs originating from the emission of excitonic complexes. Furthermore, our study investigates the twist-angle dependence of valley properties by fabricating devices with stacking angles of $\theta\sim1\degree$,  $\theta\sim4\degree$ and $\theta\sim18\degree$. Strengthened by density functional theory calculations, the results suggest that, depending on the twist angle, the conduction band minima and hybridized states at the \textbf{Q}-point promote the formation of intervalley hybrid trions involving the \textbf{Q}-and \textbf{K}-points in the conduction band and the \textbf{K}-point in the valence band. By revealing the gate control of exciton species in twisted homobilayers, our findings open new avenues for engineering multifunctional optoelectronic devices based on ultrathin semiconducting systems.


\end{abstract}

\keywords{twisted-homobilayers, MoSe$_2$, vdW heterostructures, exciton, trion, intervalley trions, electrostatic doping} 

\maketitle

\section*{Introduction}

Van der Waals heterostructures (vdWHs) based on stacked monolayers (MLs) of transition metal dichalcogenides (TMDs) have emerged as striking artificial semiconductor systems due to their unique characteristics driven by lattice mismatch effects \cite{Seyler2019, Alexeev2019, Sung2020, Andersen2021}. The wide variability of the materials, combined with the twist angle between the layers, lead to additional degrees of freedom that play a crucial role in modulating the spin-valley dynamics of van der Waals systems \cite{Seyler2019, Alexeev2019, Sung2020, Andersen2021, Villafa2023, Marcellina2021, Volmer2023} and their intra- and interlayer exciton complexes.
However, while studies of TMD heterostructures have predominantly centred on vdWHs composed of different MLs, an intriguing avenue opens up in exploring twisted homobilayers (t-BLs). Those stacked layers of the same material with a controlled rotation angle potentially exhibit more pronounced spin-valley properties than heterobilayers, in which the absence of lattice mismatch can lead to the formation of hybridized states \cite{Sung2020,Andersen2021,Villafa2023}, reconstructed domains \cite{Sung2020} and inter- and intralayer excitons trapped in moiré superlattices with a periodicity of tenths of nanometers\cite{Andersen2021, Villafa2023}. 

Exploring the electrostatic landscape of TMD-based vdWHs has also appeared as a potential avenue to fingerprint the twist-angle characteristics through exciton complexes photoluminescence response\cite{Sung2020,Andersen2021,Li2020}. However, the small oscillator strength of interlayer excitons (IX) \cite{VanderDonck2018,Zhang2020,Yu2015,Wu2018} attributed to significant electron and hole separation in momentum and real spaces\cite{Seyler2019, Alexeev2019, Sung2020, Andersen2021} has limited investigations mainly to their optical properties being hardly seen at temperatures that not cryogenics. Intralayer exciton, on the contrary, is a promising platform for studying optoelectronic properties in vdWHs, even at room temperature (RT), due to their reported large oscillator strength \cite{Mak2012,Mak2010,Chernikov2015}. Moreover, previous experimental\cite{Villafa2023, Marcellina2021} and theoretical\cite{Hagel2023} reports suggest that t-BLs present significant measurable variations in the intralayer excitons, a result weakly observed in heterobilayers due to the suppression of their intralayer optical response of at least one layered material\cite{Alexeev2019,Volmer2023}. Noteworthy, although the understanding of vdWHs semiconducting bilayers has achieved a trustworthy room throughout the years, a thoughtful understanding of the limit dynamics of exciton complexes in twisted angle-dependence till room temperature still requires further investigation.
 
In this work, we report the ability to control exciton complexes in MoSe$_2$ homobilayers through gated micro-photoluminescence ($\mu$PL) spectroscopy at RT. By varying the twist angle ($\theta$) between the MoSe$_2$ MLs, we observe an angle-rotation-dependent modulation of the charge-carrier concentration, strongly impacting neutral and charged (trions) intralayer excitons. At R-type stacking configuration of $\theta\sim1\degree$, the t-BL exhibits an intrinsic \textit{n}-type doping behaviour that efficiently enhances an exciton-to-trion conversion by electrostatic doping. Conversely, the large angle mismatch of the t-BL with $\theta\sim18\degree$ presents a close to charge-neutrality character, in which the neutral exciton features the strongest emission response independently of the applied bias. Density functional theory (DFT) calculations predict the existence of a conduction band minimum (CBM) located in the \textbf{Q}-point of the Brillouin zone instead of \textbf{K}/\textbf{K'}-points\cite{Meier2023,Miller2017} as expected for TMD MLs, in addition to hybridized states at \textbf{Q}-point. Therefore, we explain our experimental results considering that photo-excited electrons are transferred to intra- and interlayer hybrid states at\textbf{Q}- and \textbf{K}-points after photon absorption, inducing charged-excitons emission through intervalley hybrid trions formation. On the other hand, large twist angles experience no trionic enhancement, mainly due to the reduction of interlayer coupling strength effects. The findings herein provide new insights into the externally controlled optoelectronic characteristics of TMD-based vdWHs at RT, promoting the development and solid applicability of layered semiconducting material devices.

\section*{Experimental Results and Discussion}

\begin{figure*}[!htb]
    \centering
    \includegraphics[width=1\textwidth]{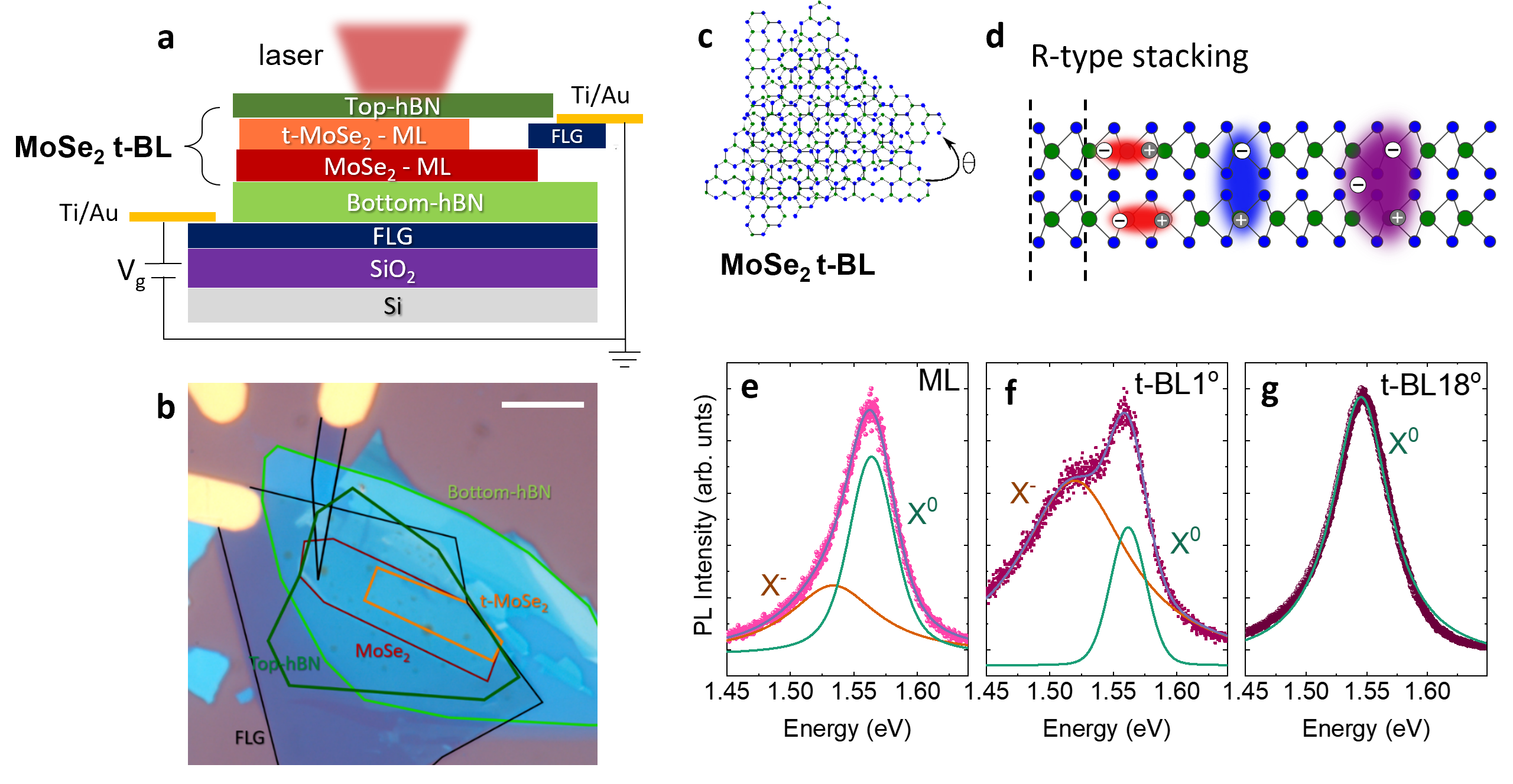}
    \caption{\small \textbf{MoSe$_2$ t-BL device properties.} (\textbf{a}), Schematic structure of the gate-controlled t-BL device. (\textbf{b}) Optical image of a fabricated t-BL device. The scale var corresponds to $10 \mu$m. (\textbf{c}) Schematic representation of the commensurate structure of a twisted MoSe$_2$/MoSe$_2$ homobilayer, (\textbf{d}) Simplified scheme of intralayer (in red) and interlayer (in blue) excitons formation in a homobilayer system. (\textbf{e-g}) PL emission of a ML, t-BL$1\degree$ and t-BL$18\degree$, respectively, at RT, measured using a CW 660~nm laser. The average excitation power used in all measurements was 10 $\mu$W. The X$^0$ and X$^{-}$ represent the neutral exciton and trion emissions, respectively. }
    \label{fig1}
\end{figure*}

We fabricated artificially stacked MoSe$_2$ homobilayers using mechanical exfoliation and dry-transfer method techniques \cite{Kim2016}, as described in the Methods section. A schematic figure of the device structure is seen in Figure~\ref{fig1}a. The experiments were performed on three devices with $\theta=(1\pm1)\degree$, $(4\pm1)\degree$ and $(18\pm1)\degree$. The twist angle deviation relies only on the equipment precision, considering that the same technique of tear-and-stick has been performed to achieve high-precision angles below $1\degree$, as reported for superconductivity in magic-angle graphene superlattices\cite{Cao2018} and studies using homobilayers\cite{Villafa2023,Sung2020,Andersen2021,Saito2020}. The Supporting Information, Fig. S1 carefully explains this twist angle-control fabrication. We decided for those specific angles to investigate the properties of small angles as well as the effects of large twist angles that still conserve R-type stacking properties (e.g., non-spin-valley locking\cite{Villafa2023}) and interlayer coupling strength\cite{Villafa2023}. 

An optical image of device $\theta\sim1\degree$ is seen in Figure~\ref{fig1}b. Both $\mu$PL and gate-dependent $\mu$PL spectra were recorded from ML and t-BL regions of every device, confirming that the MLs response of each device is similar under the same experimental conditions and, therefore, reassuring that the features observed in the t-BLs are predominantly related to the stacking configuration (e.g., different twist-angles). It is worth mentioning that in the present work, we limit the discussion to devices with $\theta\sim1\degree$ and $\sim18\degree$, since the device with $\theta\sim4\degree$ presents similar results as $\theta\sim1\degree$ (see Supporting Information, Sec. 1). 

We display in Figures~\ref{fig1}c,d, respectively, a commensurate structure of homobilayers showing a twist angle $\theta$ and a simplified scheme of intralayer (in red) and interlayer (in blue) excitons and an intervalley hybrid trion formations in hetero-and homobilayer systems. 
Figures~\ref{fig1}e--g compare the RT PL spectra at the intralayer emission range for devices with different twist angles, extracted from ML and t-BL regions under continuous-wave (CW) 660~nm laser excitation. We notice an asymmetric emission peak at the ML region (Figure~\ref{fig1}e), which we assigned under double-Voigt fitting to a convolution of two peaks, the neutral exciton (X$^0$) at $\sim$1.56~eV \cite{Tonndorf2013,Island2016} and an additional feature at lower energy ($\sim$1.52~eV) and weaker spectral weight (85\% of X$^0$). Considering the trionic binding energy between 25 to 43~meV (depending on the dielectric environment and measurement conditions) reported in the literature of TMD MLs\cite{Mak2012,Golovynskyi2021,Shang2015,Zhu2015,Drppel2017,Covre2022}, we assigned the peak at $\sim$1.52~eV to charged-exciton/trion (X$^{-}$) emission. The PL emission of t-BL$1\degree$, on the contrary, displays a comparable spectral weight between the two excitonic features X$^0$ and X$^{-}$, as seen in Figure~\ref{fig1}f, and a PL quenching of $20\times$ compared to the ML emission. This result can be ascribed to the efficient ultrafast interlayer carrier transfer between the MoSe$_2$ MLs after the dissociation of electrons and holes into separate layers\cite{Sung2020, Seyler2019, Villafa2023}. Remarkably, the t-BL$18\degree$ device (Figure~\ref{fig1}g) displays a behaviour different from the t-BL$1\degree$ case. Here, only one feature could be satisfactorily fitted at approximately 1.545 eV, indicating a major contribution from the X$^0$. Additionally, it shows no decrease in PL-yield compared to the ML, a result attributed to the substantial reduction in interlayer coupling strength between the layered MLs, at which suppression of charger transfer between layers occurs due to the large $\theta$ \cite{Villafa2023,Volmer2023}. The noticeable redshift emission of $\sim20$~meV in comparison with the ML was also reported \cite{Villafa2023}, and it can be attributed to the increased dielectric screening as a consequence of an asymmetric stacking structure formed by the adjacent MoSe$_2$ layer and the hBN layer\cite{Cho2018,Palekar2024}. Screening effects can also be responsible for the observed emission peak broadening compared with the symmetric hBN/ML/hBN response (Figure~\ref{fig1}e). 


\begin{figure*}[!htbp]
    \centering
    \includegraphics[width=0.95\textwidth]{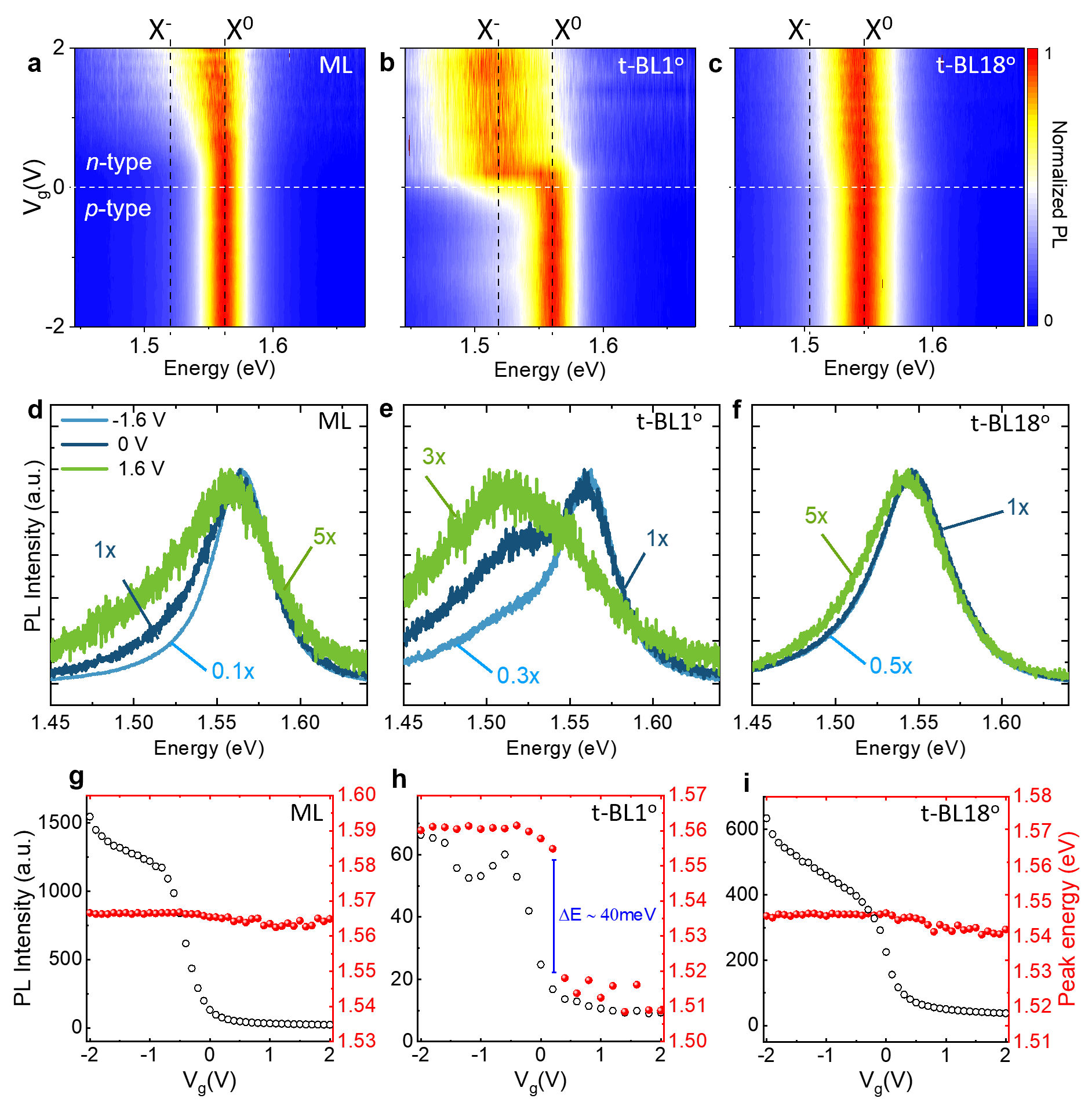}
    \caption{{\small \textbf{Gate-controlled PL modulation in MoSe$_2$ t-BLs devices.} The color plot of the PL spectra for (\textbf{a}) ML MoSe$_2$, (\textbf{b}) t-BL$1\degree$~and (\textbf{c}) t-BL$18\degree$ devices as a function of the gate voltage (V$_g$). The dashed white lines indicate the regions with \textit{p}- and \textit{n}-type doping behavior, while the expected energy position for X$^0$ and X$^{-}$ are indicated by dashed black lines. The PL intensity in the color plot was normalized for better comparison. The average excitation power used in all measurements was 10$\mu$W. (\textbf{d--f}) Selected gated-PL spectra for ML MoSe$_2$, t-BL$1\degree$~and t-BL$18\degree$ devices, respectively. The PL spectra were vertically shifted and the intensity was adjusted for better comparison. (\textbf{g--i}) Maximum PL intensity (left axis) and peak energy position (right axis) for the neutral exciton (red solid symbols) and trions (black empty symbols) in the ML MoSe$_2$, t-BL$1\degree$ and t-BL$18\degree$ devices, respectively. The ML data displayed in (\textbf{a}) were extracted from t-BL$1\degree$ device.}}
    \label{fig2}
\end{figure*}

The origin of the charge carrier concentration on t-BLs is next investigated via gate voltage-dependent PL studies in a capacitor-like geometry, as illustrated in Figure~\ref{fig1}a. Figures~\ref{fig2}a--c show colour plots of the normalized PL emission as a function of the applied back gate voltage (V$_g$) in which we notice a PL gate-dependent response with the stacking orientation (the raw PL data can be seen in the Supporting Information, Sec.~2). For the ML and t-BL$18\degree$ regions, the dominant energy remains the ascribed X$^0$ contribution ($\sim$1.56~eV and $\sim$1.54~eV in the case of ML and t-BL$18\degree$, respectively), regardless of a slight peak broadening for V$_g>0$ (Figures~\ref{fig2}a,c). In contrast, the PL emission of the t-BL$1\degree$ device shows a continuous change of energy from 1.56 to 1.52~eV together with a substantial linewidth dependence with V$_g$. 

For better comparison, Figures~\ref{fig2}d--f show the extracted gated-PL spectra for V$_g = -1.6~\text{V}$, $0~\text{V}$, and $1.6~\text{V}$ for ML, t-BL$1\degree$ and t-BL$18\degree$ devices, respectively. 
For the ML case (Figure~\ref{fig2}d), we observe an asymmetric PL emission broadening towards positive V$_g$. As expected, in this device's architecture, V$_g$ establishes the charge modulation (e.g., electrostatic doping), and hence, the Fermi level (E$_F$) shifts within the material. As such, through an applied positive (negative) bias, electrons (holes) are driven into the MoSe$_2$ active material. The emission of a MoSe$_2$ ML at RT typically consists of a dominant feature ascribed to X$^0$\cite{Tonndorf2013,Ross2013}. However, depending on the charge carrier concentration, the X$^0$/X$^-$ ratio can be altered by introducing holes or electrons in the system. Considering the results displayed in Figures~\ref{fig2}a,d, we note our crystal's intrinsic \textit{n}-type doping nature due to the trionic appearance only under positive V$_g$. Electrostatic doping effects seem to appear even more intense in the t-BL$1\degree$ device, where the overall excitonic emission is strongly tuned by just altering the charge doping, and thus, modulating what we ascribed as X$^-$ response (see Figure~\ref{fig2}e). The t-BL$18\degree$ device, on the contrary, exhibits a weak spectral weight variation (Figure~\ref{fig2}f) as an explicit characteristic of intrinsic charge neutrality. Significant attention should be paid to the electrostatic doping effects observed in the three distinct devices. On collecting the gate-voltage-dependent PL, we extracted the emission intensities and the maximum peak energy and displayed the results in Figures~\ref{fig2}g--i as a function of V$_g$. We first notice a drastic PL intensity modulation in the three regions through the applied bias. As well-stated, the electrostatic doping determines the carrier recombination process in an intrinsic \textit{n}-type semiconductor, tuning between charge neutrality for V$_{g}<0$ to high carrier concentration for V$_{g}>0$ \cite{Mak2012,Lien2019,Cadore2024,Sze2002,Plechinger2016,Ross2013}. 

Meanwhile, a high gate-induced doping level may not only provide the formation of charged excitons but promote nonradiative recombination processes\cite{Stoneham1981} (e.g., trap-assisted recombination\cite{Stoneham1981} or Auger recombination\cite{Stoneham1981,Lien2019, Mak2012, Zhu2015,Meng2020, Cadore2024}), responsible for reducing the PL yield drastically. As depicted in Figure~\ref{fig2}g, although V$_g$ sweeping alters the PL intensity, the dominant peak energy belongs to the X$^0$ emission. In contrast, the t-BL$1\degree$ device region shows a strongly gate-dependent PL under the same measurement conditions as the ML region, leading to a robust exciton-trion conversion mechanism as indicated in Figure~\ref{fig2}h by the blue vertical arrow. Similar results have been reported for several heavily doped TMD ML devices (e.g., WS$_2$ MLs \cite{Wang2017,Kwak2021,Cadore2024,Zhu2015,Shang2015}). Our findings, however, suggest that the chosen twist-angle of the t-BL plays an essential role in the charge carrier concentration of the system, as we further discuss with the help of theoretical calculations. It is worth mentioning that all t-BL devices were fabricated onto bottom hBN flakes of similar thickness (~20nm). Therefore. Therefore, the observed spectral changes cannot be associated with differenceswith differences in induced charge carriers by V$_{g}$. Between V$_g$ of $-1.6$~V and $-0.8$~V (Figure~\ref{fig2}h), the t-BL$1\degree$ device PL emission shows a dip observed over several applied-bias cycles. To this date, we could not address a specific reason that might alter the PL response of the t-BL$1\degree$ device in this V$_{g}$ range.

In Figure~\ref{fig2}, we notice that at negative V$_g$, the t-BL$1\degree$ device shows no saturation point upon the same values applied to the ML and the t-BL$1\degree$ regions. The overall scenario of the t-BL$18\degree$ device resembles the ML result (Figure~\ref{fig2}c), wherein the conversion of excitons into trions has not been observed up to the limit of our applied positive V$_g$. Such response indicates a weak interaction between the MLs, likely caused by increased interlayer distance dependent on the twist angle \cite{Volmer2023,Villafa2023,FariaJunior2023}. However, one must not overlook the potential impact of ultrafast carrier scattering between the intra-band of adjacent layers, which can also disrupt the number of electron and hole carriers recombining radiatively in the same layer, ultimately decreasing the exciton emission. We highlight that, although this work presents a careful analysis regarding the exciton complexes behaviour in TMD stacking devices, further studies may be needed to address in detail the minor changes in the energy emission behaviour. The data regarding the ML region of device t-BL$18\degree$ presents a similar response to the ML region of t-BL$1\degree$ (Figure~S6).

\begin{figure*}[!htbp]
    \centering
    \includegraphics[width=0.8\textwidth]{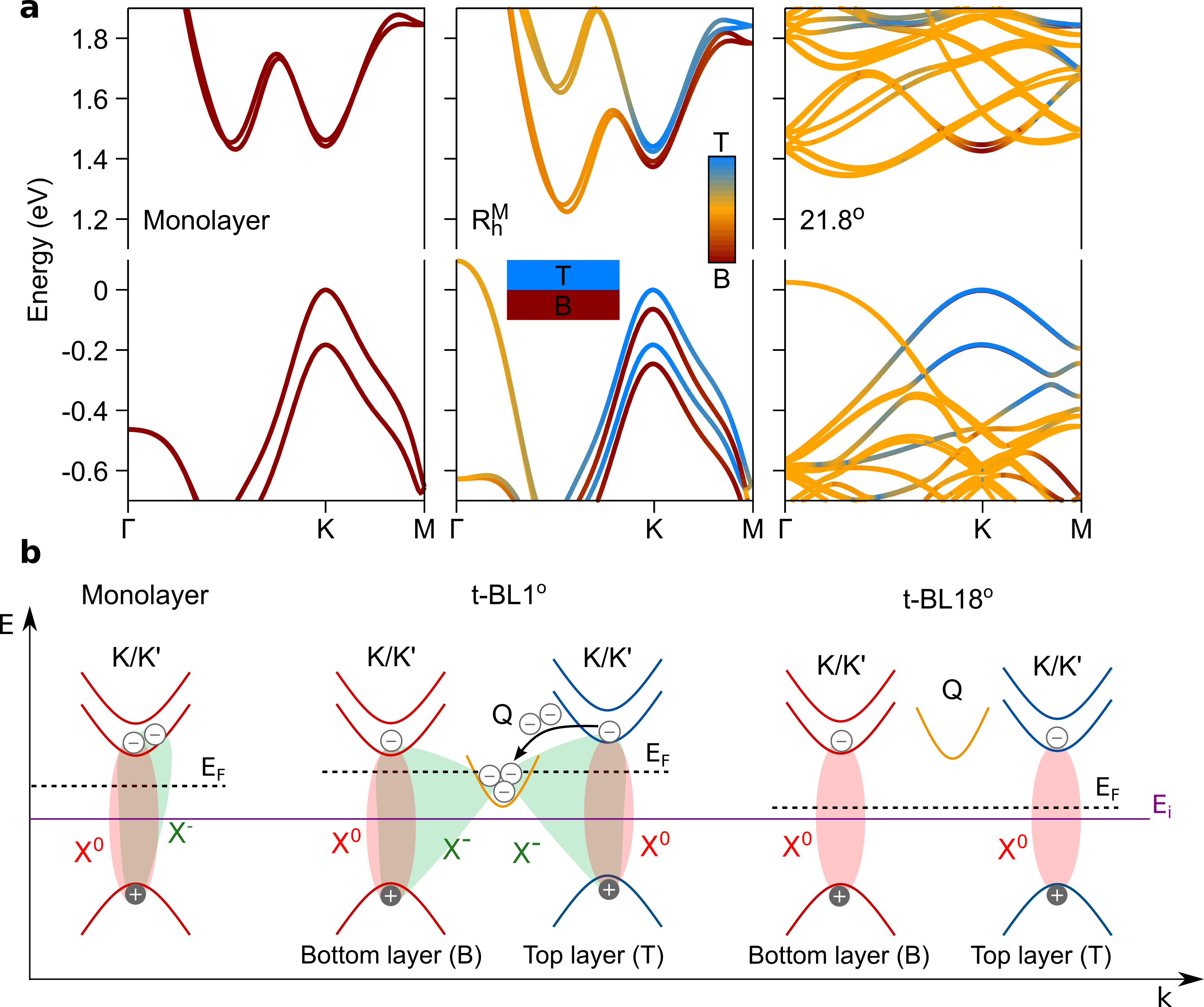}
    \caption{{\small \textbf{Band structure representation of MoSe$_2$ t-BLs devices.} (\textbf{a}) Calculated band structures for ML MoSe$_2$, R$_{\text{h}}^{\text{M}}$-stacking bilayer MoSe$_2$, and $21.8\degree$ twisted bilayer MoSe$_2$ along high-symmetry lines of the hexagonal first Brillouin zones. The color code represents the wave function localization throughout the top (T) and bottom (B) layers. For the 21.8\degree~ case, the conduction and valence bands near the \textbf{K} valleys are nearly degenerate, and the first Brillouin zone is smaller since the supercell has a lattice parameter of 8.702 $AA$, nearly 2.65 times larger than the monolayer unit cell. (\textbf{b}) Schematic picture of the band structure and exciton complexes formation of a MoSe$_2$ ML, t-BL$1\degree$ and t-BL$18\degree$ under non-applied bias, V$_{g}=0$. E$_i$, E$_F$, X$^0$, and X$^{-}$ represent the intrinsic charge neutrality level, Fermi level, neutral exciton, and trion, respectively.}}
    \label{fig3}
\end{figure*}

To gain additional insight into the optical properties of our gate-controlled t-BLs, we performed DFT calculations to reveal the twist-angle-dependent characteristics of the different systems, such as the relevant band edges, band alignment, and interlayer coupling strength \cite{FariaJunior2023}. Further details about the DFT calculations are in Supporting Information, Section 3.
The selected configurations for the homobilayer systems are based on the energetically favourable stacking configuration in the case of $\theta=0\degree$ (R$_{\text{h}}^{\text{M}}$-stacking), and the smallest structure closest to our experimental value of $\theta\sim18\degree$, which is $\theta\sim21.8\degree$ (Figure~\ref{fig3}a). 
In both configurations, the relevant band edges involved in the optical process discussed here are the CBM at \textbf{K}- and \textbf{Q}-points, and valence band maximum at \textbf{K}- and {$\mathbf{\Gamma}$}-points. Particularly, our calculations show that these different bands' relative energy separation and interlayer hybridization strongly depend on the twist angle, in line with other reports\cite{Villafa2023,Sung2020}. We acknowledge that, although our experimental results may not be sufficient to exclude any additional presence of moiré excitons at RT completely, we understand that this confinement effect may not be noticeable at room temperature since the carrier recombination is very sensitive to the environment, tending, therefore, to be less pronounced at higher temperatures \cite{Hagel2023,Wu2022,Steinhoff2024,Rossi2024,Seyler2019}. Nonetheless, one must consider that the IX formation in a twisted bilayer still plays an essential role in explaining the results presented herein. The observed intralayer PL-yield reduction of t-BL$1\degree$ is attributed to the ultrafast photo-excited carrier transfer between the adjacent MLs, induced by the proximity between layers, and the hybridization of states, thus promoting the IX recombination \cite{Villafa2023,Sung2020,Volmer2023} at low temperatures, as reported\cite{Villafa2023,Sung2020,Volmer2023}. Conversely, the carriers IX formation in t-BL$18\degree$ is expected to decrease considerably since the large stacking angles alter the interlayer distance, suppressing the interlayer transfer channels \cite{Villafa2023,Volmer2023}. However, this understanding lacks an additional comprehension of the high variation of charge carrier concentration in our systems. 

Let us now discuss the effect of the applied gate-voltage in light of the calculated band structures for the t-BL systems. The exciton manipulation through gate-voltage is explained in the case of our t-BLs considering the forming band structure after the layer stacking. Figure~\ref{fig3}b schematizes the energy level diagrams of t-BL$1\degree$ and t-BL$18\degree$ (at V$_g=0$~V), based on our experimental and theoretical inputs. For the t-BL$1\degree$ device, our theoretical analysis suggests that the CBM at the \textbf{Q}-point, which lies at the lowest energy, favours the electrons created by photon absorption and electrostatic doping to be pushed into that \textbf{Q}-point. Moreover, due to the wave function delocalization of the \textbf{Q}-point in both layers (orange colour in Figure~\ref{fig3}a), the observed trions in the twisted homobilayers are, in fact, interlayer and intervalley hybrid species, similar to those recently observed in MoSe$_2$/WSe$_2$ samples at room temperature\cite{Meier2023}, as schematized in Figure~\ref{fig3}b (t-BL$1\degree$ case). This physical picture indicates a robust intervalley trionic response in the t-BL$1\degree$, which, hence, explains the charge carrier manipulation observed through gated-PL as a signature of a highly-doping regime in semiconductors\cite{Wang2017,Kwak2021,Cadore2024,Zhu2015,Shang2015,Sze2002}. 

In contrast, the t-BL$18\degree$ gated-PL initially gives an impression that the system acts as a semiconductor closer to the neutrality point in comparison to the t-BL$1\degree$ device, in which the applied voltage barely promotes the appearance of trions in the emission response. This behaviour can be explained by examining our DFT analysis of the $21.8\degree$ case. Our calculations reveal that the conduction and valence bands at the \textbf{K}-point are nearly degenerate and strongly layer-localized, thus causing the t-BL$18\degree$ to behave as two uncoupled MLs, further supporting the resemblance of the PL spectra at RT to that of the ML case (see Figures~\ref{fig1}d,f).
The generated electrons may preferentially stay at K-point, promoting the excitonic formation mainly as an electron-hole pair at \textbf{K}-points at conduction and valence bands. 

RT gated-controlled PL supports those findings (Figures~\ref{fig2}c,f,i) by showing no relevant changes when altering the charge carrier concentration. We note that intervalley trions have been reported for TMD MLs\cite{Lyons2019,Volmer2017,YZhang2020,Pei2023}; however, mainly through external perturbation \cite{Pei2023,YZhang2020} that alter the valley or spin properties and, therefore, favour the formation of intervalley trions. Here, we have shown that the stacking angle, on the other hand, can act as a charge-carrier controller just as much as applied fields. Moreover, we emphasize that the resulting intervalley trions observed in the t-BLs exhibit a hybrid interlayer-intralayer nature, which has been only reported in the case of cryogenic temperatures of vdWHSs. Noteworthy, although the {$\mathbf{\Gamma}$}-point also presents the lowest energy point in the VB of t-BL$1\degree$ and t-BL$18\degree$, intervalley trions arising from electrons in \textbf{K}-point bounded to the holes in {$\mathbf{\Gamma}$}- and \textbf{K}-points are not expected through the range of applied V$_g$ due to the intrinsic \textit{n}-doping character of our devices. 

\section*{Conclusion}

In summary, this work investigates the optoelectronic properties of MoSe$_2$ homobilayers at room temperature through gated-PL spectroscopy. The experimental results showed an angle-alignment-dependent charge carrier concentration control, confirmed through small and large stacked-layer rotation devices. Combined with DFT calculations, our experimental results suggest that small twist angles favour the conduction band minima alteration at the \textbf{Q}-point, facilitating the formation of intervalley hybrid trions. In contrast, a large-twist-angle mismatch provided a non-doped behaviour resembling the optical response of layered-decoupled monolayers. We also demonstrate that, even at room temperature, intralayer excitons may provide a trustworthy signature for the twist-angle van der Waals heterostructures properties. Our findings significantly expanded the knowledge of TMD homobilayers by controlling their emission properties, which provides a fruitful groundwork for future potential applications of van der Waals semiconductor devices.

\section*{Methods}

\paragraph*{device preparation.} 
The mechanical exfoliation\cite{Novoselov2004} and the dry-transfer method\cite{CastellanosGomez2014} were performed through all 2D materials (MoSe$_2$ crystal, graphite~(Gr) and hexagonal boron nitride~(hBN)) used to fabricate our twisted MoSe$_2$ homobilayers. The MoSe$_2$ MLs were obtained by exfoliating TMD bulk crystals synthesized through flux zone growth \cite{Nagao2017} and identified by optical contrast and $\mu$PL spectroscopy. To create the t-BL devices, we used the tear-to-stack method \cite{Kim2016,Cao2018}. The vdWHs composed by ($\sim$10~nm)hBN/MoSe$_2$/MoSe$_2$/($\sim$20~nm)hBN was transferred onto pre-selected few-layer Gr (FLG) flakes on a 285~nm SiO$_2$/Si substrate.
The devices were gated through electrical contacts made of Au(50~nm)/Cr(5~nm) patterned by electron-beam lithography and deposited using electron-beam evaporation. 

\paragraph*{PL and gate-dependent PL measurements.} 
$\mu$PL measurements were performed at room and cryogenic (6~K) temperatures, the latter being carried out in a helium flow cryostat. In both conditions, the devices were excited non-resonantly using a CW $660$~nm laser. Devices were excited, and the emission was collected through 50x and 20x long working distance objectives (NA=0.4), for the PL and PL-gate measurements, respectively. 
For PL-gated measurements, a pulse generator in combination with a source-meter was used. The PL emission was detected using a spectrometer equipped with thermoelectrically cooled charged-coupled device (CCD) (spectral resolution: 0.12~nm). 

\section*{Supporting Information}

Non-normalized gate-dependent PL for all devices, DFT calculation method, and supporting figures.  

\section*{Conflict of interest}

The authors declare no competing interest.

\section*{Data availability statement}

The data that support the findings of this study are available from the corresponding author upon reasonable request. 

\section*{Acknowledgement}

BLTR, YY, AKS, CCP and SR  acknowledge financial support from the Deutsche Forschungsgemeinschaft (DFG, German Research Foundation) via projects Re2974/26-1 (within priority programm SPP 2244) and Re2974/21-1. PEFJ and JF acknowledge financial support of the DFG via SFB 1277 (Project-ID 314695032, projects B07 and B11), SPP 2244 (Project No. 443416183), and of the European Union Horizon 2020 Research and Innovation Program under Contract No. 881603 (Graphene Flagship). ARC acknowledges the financial support from the Brazilian Nanocarbon Institute of Science and Technology (INCT/Nanocarbono) and the National Council for Scientific and Technological Development (CNPq).
The authors acknowledge Prof. Dr. Leandro Malard for enlightening discussion.

\bibliographystyle{unsrt} 
\bibliography{reference}

\end{document}